\documentstyle[psfig]{mn}
\def\la{Ly$\alpha$}
\def\etal{et~al.}
\def\spose#1{\hbox to 0pt{#1\hss}}
\def\lta{\mathrel{\spose{\lower 3pt\hbox{$\mathchar"218$}}
     \raise 2.0pt\hbox{$\mathchar"13C$}}}
\def\gta{\mathrel{\spose{\lower 3pt\hbox{$\mathchar"218$}}
     \raise 2.0pt\hbox{$\mathchar"13E$}}}
\def\Ha{H$\alpha$}

\def\iraf{{\sc iraf}}
\def\noao{{\sc noao}}

\def\Ho50{$H_0 = 50$km\,s$^{-1}$\,Mpc$^{-1}$}

\title{More redshifts of powerful equatorial radio sources from the BRL
sample}

\author[P.~N.~Best \etal]{P.~N.~Best,$^1$\thanks{Email:
pbest@strw.leidenuniv.nl} H.~J.~A.~R\"ottgering$^1$ and
M. D. Lehnert$^2$\\ 
$^1$ Sterrewacht Leiden, Postbus 9513, 2300 RA Leiden, the Netherlands\\ 
$^2$ Max-Planck-Institut f{\"u}r extraterrestrische Physik, P.O.Box 1603,
87450 Garching, Germany}

\pagerange{\pageref{firstpage}--\pageref{lastpage}}
\pubyear{1999}
\begin{document}
\label{firstpage}
\setcounter{topnumber}{3}
\setcounter{dbltopnumber}{4}

\maketitle

\begin{abstract}
\noindent A new sample of very powerful radio sources, defined from the
Molonglo Reference Catalogue, was recently compiled by Best,
R{\"o}ttgering and Lehnert (1999). These authors provided redshifts for
174 of the 178 objects in the sample, making the sample 98\%
spectroscopically complete. Here, redshifts for three of the remaining
galaxies are presented, confirming the optical identifications and raising
the spectroscopic completeness of the sample to 99.5\%; only 1059$-$010
(3C249) is currently without redshift.
\end{abstract}


\section{Introduction}

Radio sources have many important roles to play in astrophysical and
cosmological studies (e.g. see McCarthy 1993 for a review)\nocite{mcc93}.
In order to provide a large, spectroscopically complete sample of luminous
radio sources accessible to both northern radio interferometers such as
the Very Large Array (VLA) and to large southern telescope facilities,
such as the Very Large Telescope, Gemini South, and the Atacama Large
Millimetre Array (ALMA), Best \etal\ \shortcite{bes99e} recently defined a
new sample of very powerful equatorial radio sources from the Molonglo
Reference Catalogue (MRC; Large \etal\ 1981)\nocite{lar81}, according to
the criteria (see Best \etal\ for details): $S_{\rm 408 MHz} \ge 5$\,Jy,
$-30^{\circ} \le \delta \le +10^{\circ}$, $|b| \ge 10^{\circ}$. This
sample (hereafter the BRL sample) consists of 178 objects and, following
radio imaging, optical imaging, and spectroscopic observations,
spectroscopic redshifts were provided for 174 of these in the original
paper. The host galaxies of the remaining four sources were all optically
identified, but no spectroscopic redshifts were obtained.

In this paper, spectroscopic redshifts are derived from new observations
of three of these remaining four objects, 1413$-$215, 1859$-$235 and
1953$-$077 (3C404). In Section~\ref{obs}, details of the observations and
data reduction are provided. The reduced spectra are presented and
discussed in Section~\ref{results}. The reader is referred to Best \etal\
\shortcite{bes99e} for a complete description of the sample and its
properties.
\vspace*{-4mm}

\section{Observations and Data Reduction}
\label{obs}
 
Long--slit spectra of 1859$-$235 and 1953$-$077 were taken using the
duel--beam ISIS spectrograph on the William Herschel Telescope (WHT) in
photometric conditions during service time on the night of 1999 July 5
(see Table~1 for details). The observations were made using the 5700\AA\
dichroic and the R158B and R158R gratings in the blue and red arms of the
spectrograph. In the blue arm this provided a spatial scale of 0.19 arcsec
per pixel and a spectral resolution of about 19\AA, and in the red arm a
spatial scale of 0.36 arcsec per pixel and a spectral resolution of about
12\AA.

The data were reduced using standard packages within the \iraf\ \noao\
reduction software. After subtraction of the bias level, the spectroscopic
data were flat--fielded using observations of internal calibration lamps,
and the sky background was removed. The two exposures of each galaxy were
combined, removing cosmic ray events, and one dimensional spectra were
extracted from an angular extent of 2.9 arcsec along the slit. The
extracted spectra were wavelength calibrated using observations of CuNe
and CuAr arc lamps, and flux calibration was achieved using observations
of the spectrophotometric standard star Kopff 27. The determined fluxes
were corrected for any atmospheric extinction arising from the non--unity
airmass of the observations.

\addtocounter{table}{1}
\begin{table*}
\caption{\label{emistab} Emission line properties of the galaxies. In
addition to the derived galaxy redshifts, the integrated fluxes of the
emission lines are given, together with their deconvolved velocity
full--width at half--maxima and their rest--frame equivalent widths. These
properties were calculated in the manner described in Best \etal\ (1999).
}

\begin{tabular}{lclrrrrrr}
~~~Source & Redshift & Line & 
\multicolumn{2}{c}{Line Flux} & 
\multicolumn{2}{c}{FWHM (deconv.)} & \multicolumn{2}{c}{Eq. Width} \\
       &                            &      & 
\multicolumn{2}{c}{[10$^{-16}$\,erg/s/cm$^2$]} &
\multicolumn{2}{c}{[km\,s$^{-1}$]} & \multicolumn{2}{c}{[\AA]}\vspace*{2mm} \\
{\bf 1413$-$215}&$1.116 \pm 0.001$&[NeV]  &1.0&$\pm$ 0.1\hspace*{6mm} &
$<390$ &   &33 &$\pm$ 5\\
                &                 &[OII]  &\hspace*{4mm}2.5&0.3\hspace*{6mm} &
760&$\pm$ 280\hspace*{2mm} &99&19\\
                &                 &[NeIII]&0.6&      0.2\hspace*{6mm} &
\hspace*{4mm} 730   &590\hspace*{2mm} &18 &6\vspace*{2mm}\\
{\bf 1859$-$235}&$1.430 \pm 0.001$&[CIV]  &3.4&      0.6\hspace*{6mm} &
$<1300$&   &103&5\\
                &                 &HeII   &4.6&      0.6\hspace*{6mm} &
 1230  &800\hspace*{2mm} &~~173&6\\
                &                 &CIII]  &4.7&      0.5\hspace*{6mm} &
 800   &410\hspace*{2mm} & 75&14\\
                &                 &CII]   &6.7&      1.1\hspace*{6mm} &
$<1950$&   &142&4\\
                &                 &MgII   &1.8&      0.3\hspace*{6mm} &
 1180  &480\hspace*{2mm} & 24&4\vspace*{2mm}\\
{\bf 1953$-$077}&$1.338 \pm 0.002$&[OII]  &7.6&      1.2\hspace*{6mm} &
  880  &410\hspace*{2mm} &195&10\\
\end{tabular}
\end{table*}
\addtocounter{table}{-2}

\begin{figure*}
\begin{tabular}{lr}
\psfig{file=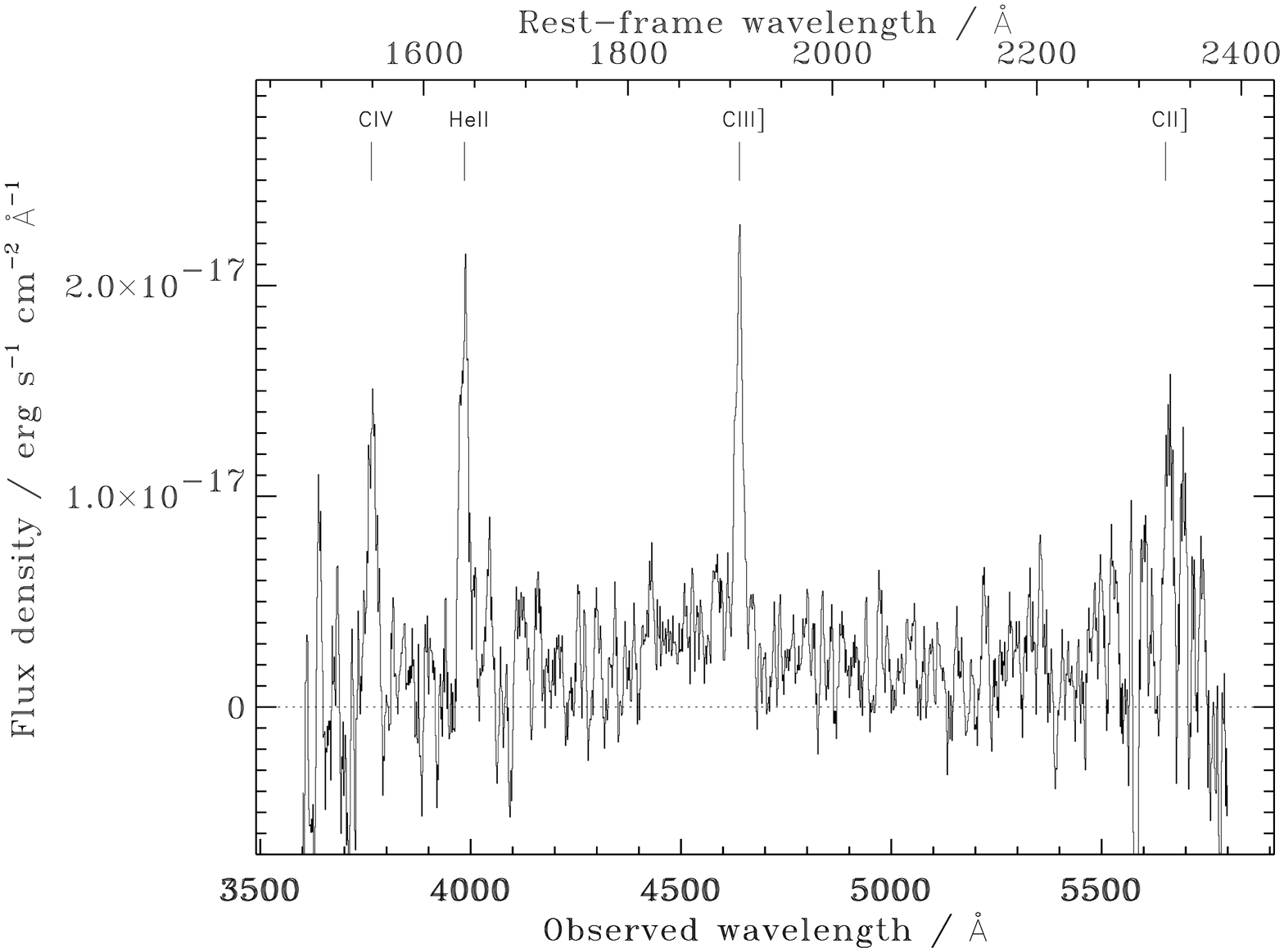,angle=90,width=7.8cm,clip=}
&
\psfig{file=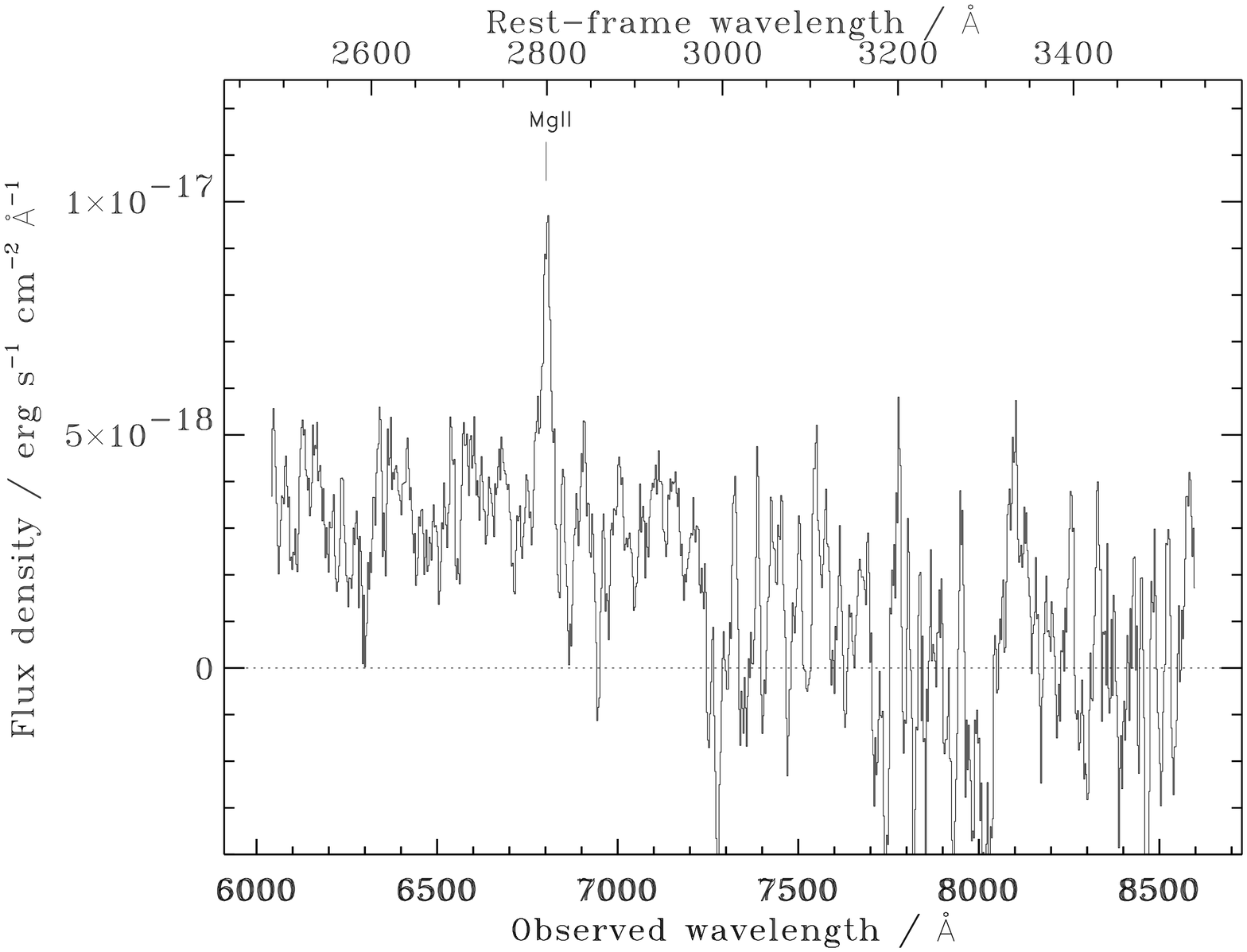,angle=90,width=7.8cm,clip=}
\end{tabular}
\caption{\label{1859rfig} The WHT ISIS blue arm (left) and red arm (right)
spectra of {\bf 1859$-$235}.}
\end{figure*}

1413$-$215 was observed at the Keck II telescope in photometric conditions
during evening twilight on 1999 July 11 (see Table~1). The observations
were made using the Low--Resolution Imaging Spectrograph (LRIS; Oke \etal\
1995)\nocite{oke95} with the 150 line\,/\,mm grating (7500\AA\ blaze),
providing a spatial pixel scale of 0.21 arcsec and a spectral resolution of
about 25\AA. The galaxy was shifted 10$''$ along the slit between two
separate observations to reduce fringing effects. Data reduction followed
essentially the same procedure as outlined for the WHT observations,
except that the spectrum was extracted from an angular extent of 2.1$''$
along the slit (due to a smaller spatial extent of the object). Feige 110
and HZ44 were used for flux calibration.

\begin{figure}
\centerline{
\psfig{file=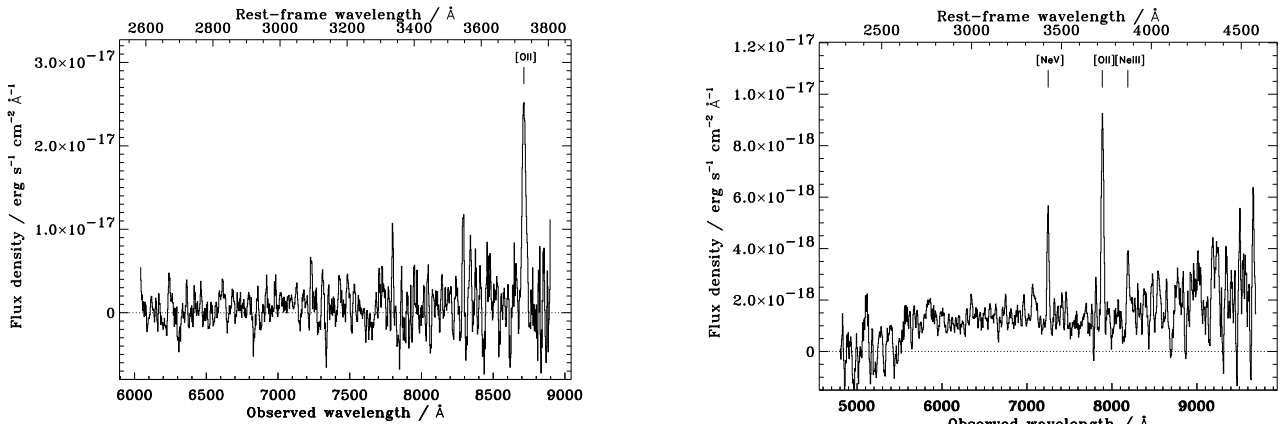,angle=90,width=7.8cm,clip=}
}
\caption{\label{1953fig} The WHT ISIS red arm spectrum of {\bf 1953$-$077}.}
\end{figure}

\begin{figure}
\centerline{
\psfig{file=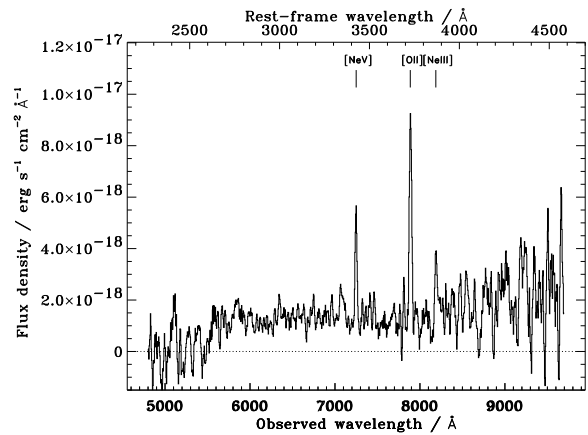,angle=90,width=7.8cm,clip=}
}
\caption{\label{1413fig} The Keck LRIS spectrum of {\bf 1413$-$215}.}
\end{figure}

\section{Results and Discussion}
\label{results}

The extracted spectra of the three galaxies are provided in Figures~1, 2
and~3, and details of the emission line properties are provided in
Table~\ref{emistab}. Emission lines are detected for all three objects,
confirming the identifications proposed in the paper by Best \etal\
\shortcite{bes99e}.  For 1859$-$235 and 1413$-$215, several emission lines
are detected, providing unambiguous redshift measurements. For 1953$-$077
only a single strong emission line is detected, at 8714\AA. This emission
line is assumed to be [OII]~3727 for a number of reasons: (i) were this
[OIII]~5007 or \Ha\ (or any other weaker line), then the lack of any other
strong emission lines between 3500 and 9000\AA\ would be very surprising;
(ii) weak continuum emission is detected in the red--arm observations down
to about 6000\AA, ruling out the possibility that the line is \la; (iii)
if the line is [OII]~3727 then, given the $R$ magnitude of the source ($R
= 22.90$), the derived redshift places it in the middle of the $R-z$
diagram of the other radio galaxies (cf. Figure~51 of Best \etal\
1999). It appears fairly secure, therefore, that this emission line is
[OII]~3727.

\begin{table}
\caption{\label{obstab} Details of the spectroscopic observations.}
\begin{tabular}{lccccr}
~~~Source & Observ. & Tele. & Exp.         & Slit  & Slit \\
          & Date    &       & Time         & Width & PA   \\
          &         &       &  [s]         & [$''$]& [deg]\\
1413$-$215& 11/07/99& Keck  &$2 \times 600$& 1.5   & 350  \\ 
1859$-$235& 05/07/99& WHT   &$2 \times 900$& 2.5   & 90   \\
1953$-$077& 05/07/99& WHT   &$2 \times 900$& 2.5   & 340  \\    
\end{tabular}
\end{table}

Following these results, the BRL sample is now 99.5\% complete. The only
object without a spectroscopic redshift in the sample is 1059$-$010
(3C249), whose very faint $R$ magnitude ($R = 24.20$; Best \etal\ 1999)
suggests a minimum redshift of 1.5. Spinrad, Stern and Dey (private
communication) have attempted, without success, to obtain a redshift for
this object using the Keck Telescope, and Rawlings (private communication)
has carried out near--infrared spectroscopy with UKIRT in the J--band and
between 1.6 and 2.2 microns, detecting the continuum but no
lines. Obtaining the final redshift in the sample may prove difficult.

\section*{Acknowledgements} 

This work was supported in part by the Formation and Evolution of Galaxies
network set up by the European Commission under contract ERB FMRX--
CT96--086 of its TMR programme. The WHT is operated on the island of La
Palma by the Isaac Newton Group in the Spanish Observatorio del Roches de
los Muchachos of the Instituto de Astrofisica de Canarias. The W. M. Keck
Observatory is operated as a scientific partnership among the University
of California, the California Institute of Technology, and NASA. The
Observatory was made possible by the generous financial support of the
W. M. Keck Foundation. The authors are grateful to the WHT service
observer, John Telting, to Wil van Breugel, Carlos De Breuck, Dan
Stern and Adam Stanford for kindly observing 1413$-$215 at the Keck
Telescope, and to the referee, Steve Rawlings, for helpful comments.

\bibliography{pnb} 
\bibliographystyle{mn} 
\label{lastpage}

\end{document}